\documentclass[journal]{IEEEtran}
\usepackage{cite}

\ifCLASSINFOpdf
  \usepackage[pdftex]{graphicx}
  \DeclareGraphicsExtensions{.pdf,.jpeg,.png}
\else
\fi
\usepackage{amsmath}
\usepackage{array}

\usepackage{comment}
\usepackage{color,soul}

\usepackage[switch]{lineno}

\begin{document}

%
\title{FastPET: Near Real-Time PET Reconstruction from Histo-Images Using a Neural Network}
%
%
%

\author{William~Whiteley,~\IEEEmembership{Member,~IEEE,}
        Vladimir~Panin,~\IEEEmembership{Member,~IEEE,}
        Chuanyu~Zhou,
        Jorge~Cabello,~\IEEEmembership{Member,~IEEE,}
        Deepak~Bharkhada,~\IEEEmembership{Member,~IEEE,}
        and~Jens~Gregor
\thanks{W. Whiteley and J. Gregor are with with the Department
of Electrical Engineering and Computer Science, The University of Tennessee, Knoxville,
TN, 37996 USA e-mail: william.whiteley@siemens-healthineers.com.}
\thanks{W. Whiteley,  V. Panin, C. Zhou, J. Cabello, and D. Bharkhada are with Siemens Medical Solutions Inc. USA, Knoxville, TN 37932 }
\thanks{Manuscript received Jan 30, 2020}}

%
%

\markboth{Preprint submitted IEEE Transactions on Radiation \& Plasma Medical Sciences }
{Whiteley \MakeLowercase{\textit{et al.}}: FastPET: Near Real-Time PET Reconstruction from Histo-Images Using a Neural Network}
%



\maketitle

\begin{abstract}
Direct reconstruction of positron emission tomography (PET) data using deep neural networks is a growing field of research. Initial results are promising, but often the networks are complex, memory utilization inefficient, produce relatively small 2D image slices (e.g., 128x128), and low count rate reconstructions are of varying quality. This paper proposes FastPET, a novel direct reconstruction convolutional neural network that is architecturally simple, memory space efficient, works for non-trivial 3D image volumes and is capable of processing a wide spectrum of PET data including low-dose and multi-tracer applications. FastPET uniquely operates on a histo-image (i.e., image-space) representation of the raw data enabling it to reconstruct 3D image volumes 67x faster than Ordered subsets Expectation Maximization (OSEM). We detail the FastPET method trained on whole-body and low-dose whole-body data sets and explore qualitative and quantitative aspects of reconstructed images from clinical and phantom studies. Additionally, we explore the application of FastPET on a neurology data set containing multiple different tracers. The results show that not only are the reconstructions very fast, but the images are high quality and lower noise than iterative reconstructions. 
\end{abstract}


%
\IEEEpeerreviewmaketitle

\section{Introduction}

\IEEEPARstart{P}{ositron} emission tomography (PET) is a functional imaging modality utilizing biological radioactive tracers with wide ranging applications in oncology, cardiology, neurology and medical research. During a PET scan the distribution of a radioactive pharmaceutical administered to the patient is recovered through the process of image reconstruction. Typically, the data is relatively sparse and noisy making this process a challenging inverse problem that is conventionally solved using either analytical or iterative reconstruction techniques. Analytical algorithms have a closed-form solution, are fast and produce images that are quantitatively accurate, but suffer from low visual image quality with their characteristic streak artifacts. Iterative algorithms incorporate well-studied statistical models in the reconstruction process which leads to higher visual image quality, only at a higher computational cost leading to longer reconstruction times. More recently, reconstruction algorithms have emerged that utilize deep neural networks. This new category contains hybrid techniques that combine conventional reconstruction with machine learning as well as direct reconstruction techniques where neural networks operate directly on raw data to generate images. In this paper, we explore a novel neural network for direct reconstruction with the development of FastPET, which is capable of producing image volumes in near-real time. We examine reconstruction speed and perform quantitative and qualitative image comparisons the clinical benchmark reconstruction algorithm Ordered Subsets Expectation Maximization plus Point Spread Function (OSEM+PSF) \cite{hudson:97, Panin:06}

Conventional reconstruction methods apply explicit corrections for attenuation, scatter, randoms, and other effects. In contrast, direct neural network methods learn nearly all aspects of reconstruction using the data driven method of supervised learning. A concern often associated with neural networks is their black-box nature compared to the well understood and hand-crafted conventional algorithms. This trade-off must be weighed when considering the benefits of deep learning based reconstruction. 

FastPET and other direct neural network reconstruction methods offer benefits not found with traditional or even hybrid reconstruction techniques. Most immediate is the computational efficiency. While neural networks may take days or even weeks to learn a reconstruction model, a trained neural network can produce 3D image volumes in near real-time. This can improve workflow and remove reconstruction time as a consideration when developing PET scan protocols and selecting reconstruction parameters. Notable applications that would benefit therefrom include dynamic and gated studies that often include a large number of reconstructions as well as interventional PET imaging which requires a real-time response. 

\begin{figure*}[t]
	\centering
	\includegraphics[width=\textwidth]{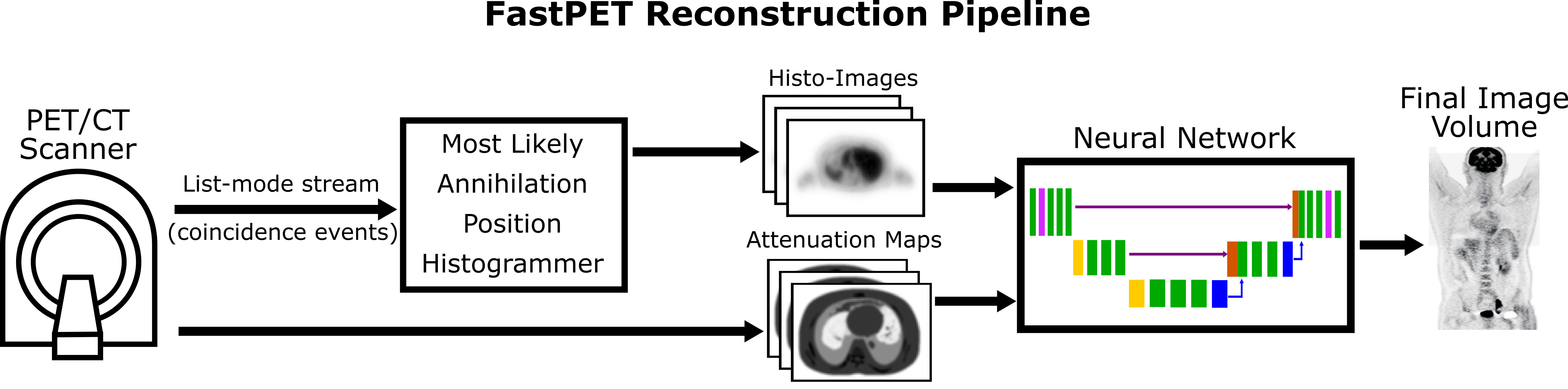}
	   \caption{The FastPET reconstruction pipeline consists of a most likely annihilation position histogrammer that places photon coincidences into a histo-image representation followed by a convolutional neural network that uses the histo-images and attenuation maps to reconstruct image volumes.}
	\label{pipeline}
\end{figure*}

Beyond computational efficiency, data driven approaches are inherently flexible in their ability to learn an underlying model provided sufficient training examples and network capacity. We show that FastPET is an example of this by demonstrating a learned mapping from noisy measurement data to a quality image with the same network on three separate data sets with no re-engineering required. Given the ever increasing availability of more powerful computational resources combined with continued improvements in deep learning methods, it is quite possible that learned reconstruction methods eventually will become common in medical imaging.

FastPET, as shown in Figure~\ref{pipeline} and described in more detail in Section \ref{method}, consists of a so-called most likely annihilation position (MLAP) histogrammer, and a convolutional neural network. Raw coincidence events are converted to histo-images \cite{Matej:09} which are fed to the neural network along with attenuation maps to create a quantitative image. This approach takes advantage of the improved timing resolution of modern PET scanners, which has reduced the position uncertainty of annihilation events, combined with neural network research on denoising and deblurring. 

Histo-images and their precursor histo-projections \cite{Vandenberghe_2006} have been studied since the development of time-of-flight (TOF) PET scanners \cite{Snyder:81}. A histo-projection is a blurred version of the final image along the TOF dimension. When histo-projections for multiple views are combined, the result is a histo-image \cite{Matej:09}. Given the timing resolution of modern PET scanners, this mapped version of the raw measurement data is exceptionally well suited for convolutional neural networks due to being locally correlated and essentially making reconstruction an image-to-image operation. 

The histo-projection approach typically exploits TOF data angular compression, such as transaxial mashing \cite{Politte:86} and axial spanning \cite{Mullani:82} where adjacent angular histo-projections are combined. TOF angular compression results in relatively sparse angular sampling requiring TOF resolution to preserve spatial resolution \cite{Vandenberghe_2006,Politte:86}. The DIRECT framework \cite{Matej:09} used a similar compression tactic with histo-images and was able to achieve faster reconstructions. The method has demonstrated potential in motion detection \cite{Ren:14} and motion pre-correction \cite{Panin:15}, where a deblurring procedure may not be necessary.

Neural networks designed for deblurring and denoising are a frequent theme of deep learning research in general \cite{jain:08,xie:12,Zhang:17,Chunwei:19} and in medical imaging in particular. Neural networks have been applied to improve the image quality in low dose \cite{Yang:17,Kang:18,Ansari:2018,Kadimesetty:19} and limited angle\cite{Han:18, lee:19} X-ray computed tomography (CT) applications. They have also been applied in magnetic resonance imaging (MRI) to remove Rician noise \cite{Jiang2017} and to reduce Gibbs artifacts \cite{Wang:17}. In the PET low dose imaging domain, they have been used to synthesize normal dose equivalents \cite{Xu:2017,Gong:18,Kaplan:2018,Lu:2019, luis:20} utilizing U-Net \cite{Ronneberger:15} or ResNet \cite{He:15} style networks. Jiao et al. \cite{Jiao:2017} presented another approach where images with the characteristic streak artifacts obtained from a simple inverse Radon transform were mitigated to the image quality standard of iterative reconstruction methods. The FastPET architecture capitalizes on some of this previous work to craft a neural network capable of deblurring  histo-images while correcting for attenuation, scatter, randoms and other effects associated with the photon detection process. 

Despite the FastPET neural network operating entirely in image space, it has more in common with direct neural network reconstruction methods than with the image space filters discussed in the previous paragraph. FastPET operates on a histogrammed version of the raw PET data like other direct neural network reconstruction methods, but in this case the histogramming of coincidence events is into image space versus the typical process of histogramming events into a scanner's geometrically defined sinogram space. Operating from sinogram space usually requires neural networks to perform an expensive memory-space operation to transform the data into the image domain. 

Generally speaking, direct reconstruction using neural networks in molecular imaging is not a new topic with active research dating back to at least the early 1990's utilizing networks of fully connected multilayer perceptrons \cite{Floyd:91,Bevilacqua:00,Paschalis:04,Argyrou:12}. More recently, artificial intelligence has become a new frontier for image reconstruction \cite{Wang:18} capitalizing on two contributing factors: 1)  growth of computational resources, especially in graphical processing units (GPUs), multi-core central processing units (CPUs) and cloud computing; and 2)  development of better optimization algorithms and widely available deep learning software libraries such as PyTorch \cite{paszke:2017} and Tensorflow \cite{tensorflow2015-whitepaper}. 

The AUTOMAP network \cite{Zhu:2018} was the first direct reconstruction neural network to use modern deep learning techniques utilizing multiple fully-connected layers followed by a sparse convolutional encoder-decoder to learn a mapping manifold directly from measurement to image space. AUTOMAP was shown capable of learning a general solution to the reconstruction inverse problem, but this generality was achieved by requiring an excessively high number of parameters limiting its application to relatively small single-slice 2D images (e.g., 128x128). DirectPET \cite{whiteley:2019} improved on this architecture for PET imaging by adding a convolutional encoder to compress the sinogram input along with a specially designed Radon inversion layer consisting of many small fully connected networks to more efficiently perform the domain transformation into image space. This allowed DirectPET to produce full size multi-slice image volumes (e.g., 16x400x400) directly from sinograms. DeepPET\cite{HAGGSTROM:2019} is another example of direct neural network reconstruction, but takes an alternative approach foregoing the memory intensive fully-connected layers and instead utilizing only convolutional layers to encode the sinogram input into a higher dimensional feature vector representation, which is then decoded by convolutional layers to produce a small 2D single-slice image (e.g., 128x128). DeepPET produces smooth images with low noise, but fails at recovering detailed structures especially at low count densities where it often becomes unstable generating erroneous images. 

As summarized in Table~\ref{network_comparison}, FastPET exhibits computational advantages compared with these other neural network approaches.  Operating entirely on histo-images allows FastPET to utilize a relatively simple network architecture based on a relatively small number of network parameters.  Unlike AUTOMAP or DeepPET, which both produce small 2D image slices, and DirectPET, which produces a multi-slice 3D image volume, FastPET is capable of generating a true 3D image volume of a size commonly found in clinical and research environments.  The axial depth, here listed as 96 slices, depends only on the amount of GPU memory available.

\begin{table}[h]
\caption{Compared with modern direct neural network reconstruction methods, FastPET is capable of full-sized 3D reconstruction while maintaining a simple architecture with the fewest network parameters. }
\label{network_comparison}
\begin{tabular}{|c|c|c|c|}
\hline 
\textbf{Method} & \textbf{Parameters} & \textbf{Network Complexity}  & \textbf{Image size} \\  \hline 
AUTOMAP   &  $1,340,000,000$          & Semi-complex  & 1x128x128   \\ \hline 
DirectPET &  $352,000,000$          & Complex & 16x400x400              \\ \hline
DeepPET   &  $65,000,000$          & Simple  & 1x128x128   \\ \hline
FastPET   &  $20,000,000$         & Simple  & 96x440x440 \\  \hline                                                              
\end{tabular}
\end{table}

\section{Method}

\label{method}

\subsection{FastPET Reconstruction Architecture}
Figure~\ref{pipeline} illustrates the FastPET pipeline starting with the PET/CT scanner generating raw data in the form of PET list-mode events and CT based attenuation maps, followed by the conversion of list-mode data into histo-images by the MLAP histogrammer, and then the neural network based image reconstruction using the histo-images and attenuation maps as input.   

The MLAP histogramming and forward pass of the neural network are both computationally efficient operations allowing practical implementations where 3D PET volumes are reconstructed in near real-time. The stream of PET list-mode data coming from the scanner is stored in a first-in-first-out (FIFO) buffer. In our experiments, the MLAP histogrammer was capable of processing this data at a rate of 5 million events per second in a single thread and about 40 million events per second using 18 threads. The histo-images can be stored either as independent dynamic frames of the duration needed to facilitate an application such as motion correction. Alternatively, a single histo-image is created by continually adding the individual list-mode events. Either way, the neural network can reconstruct the histo-images in a fraction of a second as further discussed in the section on reconstruction speed.  


\subsection{Most Likely Annihilation Position Histogrammer}
The MLAP histogrammer provides fast approximate image generation capable of producing thousands of dynamic image frames per second using standard computer hardware. The arrival time difference for two coincident events is used to estimate the most likely voxel along the line of response (LOR) where the annihilation event took place. Specifically, let the two detectors for the LOR be located at $\vec{P_1}$ and $\vec{P_2}$ and assume that the arrival time difference for the two photons is ${\Delta}t=t_1-t_2$. The MLAP is then given by
\begin{equation}\label{eq_x_mla}
\vec{P}_\text{MLAP}=\frac{\vec{P}_1+\vec{P}_2}{2}+ c \frac{{\Delta}t}{2}\frac{\vec{P}_1-\vec{P}_2}{\|\vec{P}_1-\vec{P}_2\|}
\end{equation}
where $c$ denotes the speed of light. 
The current MLAP histogrammer increments a counter for the nearest voxel, if the event was a prompt; conversely, the counter is decremented, if the event was a delay (random). The count is also scaled based on the normalization crystal efficiencies obtained from the scanner's daily quality control procedure. 
Apart from randoms subtraction and normalization, no other physical corrections such as attenuation or scatter estimation, are involved in histo-image formation.

\subsection{Neural Network Architecture}

\begin{figure*}[htbp]
	\centering
	\includegraphics[width=\textwidth]{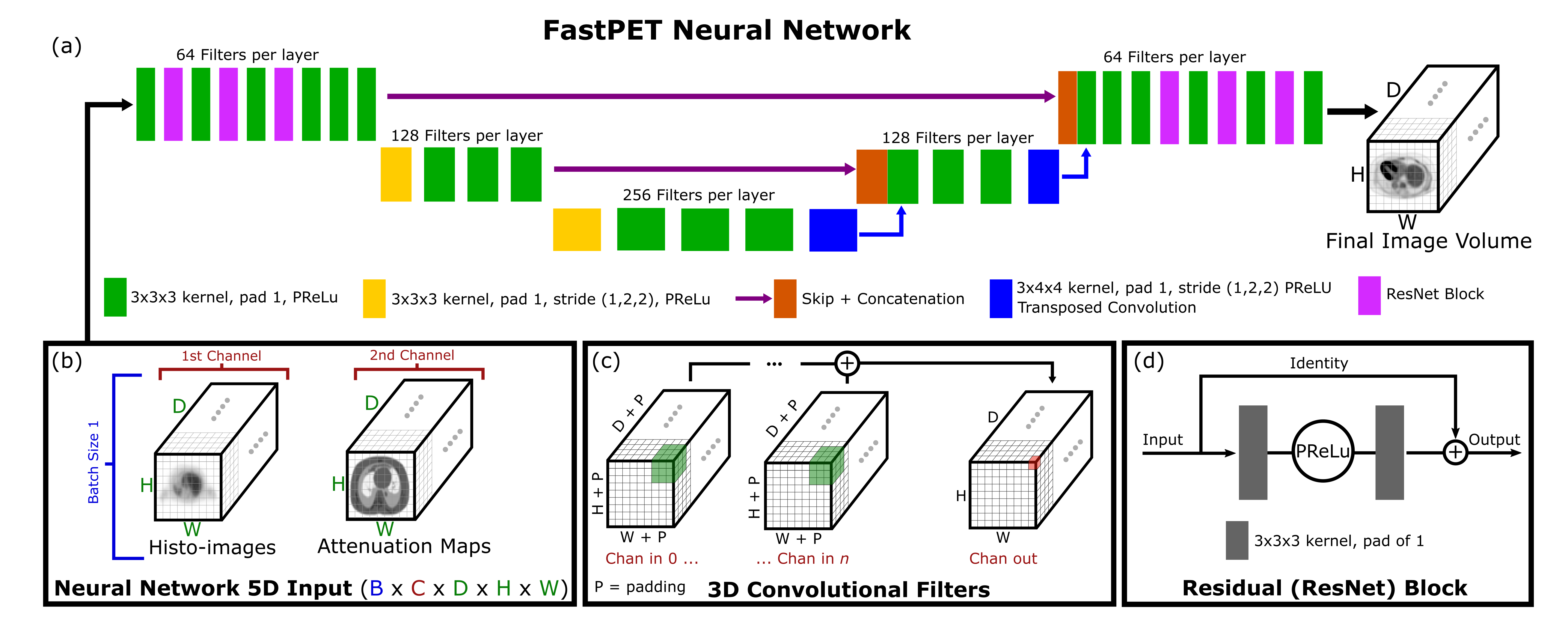}
	   \caption{The FastPET neural network utilizes a U-NET style 3D convolutional architecture with a total of 31 network layers. A kernel stride of 2 is used in the transaxial directions for down-sampling and fractionally strided transposed convolution is used for up-sampling. All activation functions are parametric rectified linear units.}
	\label{neural_network}
\end{figure*}

The FastPET neural network, shown in Figure~\ref{neural_network}(a), utilizes a U-NET \cite{Ronneberger:15} style architecture with the well-known contracting, bottle neck and expanding segments where the number of convolutional kernels is doubled each time the transaxial spatial resolution is reduced and the number of convolutional kernels is halved each time the spatial resolution is increased. This style of network also contains skip connections where the features extracted in the last layer at each spatial resolution on the contracting side are concatenated to the first layer at the same resolution on the expanding side of the network. The proposed network also includes residual blocks \cite{He:15} as shown in Figure~\ref{neural_network}(d) near the input and output that help the error gradients propagate to the earliest layers and increase training efficiency and stability. Overall, the proposed network contains 31 layers and has approximately 20 million learnable parameters. Each layer, except for upsampling, uses $3\times 3\times 3$ convolutional kernels and parametric rectified linear unit (PReLU) \cite{he:15:prelu} activation functions. Spatial down-sampling is accomplished by using a kernel stride of 2 but only in the height and width dimension (i.e. transaxial) leaving the tomographic depth constant throughout the neural network. The up-sampling is accomplished using a fractionally-strided transposed convolution with a scaling factor of 2, but again only in the height and width dimensions.

The neural network input contains batches of 3D histo-images and matching attenuation maps each with a size of $d\times h\times w$. This creates a 5D input, i.e., $b\times c\times d\times h\times w$, to the neural network as shown in Figure~\ref{neural_network}(b). The batch dimension $b$ is adjusted based on available memory and is used to simultaneously sample and reconstruct multiple volumes and update the network based on the average error gradient. The channel dimension $c$ at the input is 2 representing the histo-image and attenuation map volumes. However, after the first layer the channel dimension represents the number of convolutional filters of a given layer. 

The FastPET neural network utilizes 3D convolutional filters as illustrated in Figure~\ref{neural_network}(c). The input to a convolutional layer is padded by zero valued voxels to maintain the same data dimensions at the output. As the name implies, 3D convolutional filters move in all three dimensions, but also operate in parallel across each of the layer's channels. In a given layer with $c$ input channels, this leads to a convolutional filter of size $c\times d\times h\times w$; in our proposed network specifically, $c\times 3\times 3\times 3$ except in the case of the up-sampling layers that use $c\times 3\times 4\times 4$ kernels.  For each convolutional filter in a given layer, an output channel is generated.    

\subsection{Training and Evaluation Data Sets}
Data was acquired on a Siemens Biograph Vision 8 ring PET/CT scanner\cite{sluis:19} containing 3.2 mm$^2$ crystals with a TOF resolution of 214 ps. The target images were generated using the factory attenuation corrected TOF ordered subsets expectation maximization reconstruction protocol with point spread function correction (OSEM+PSF).  The chosen image size was $440\times440$ resulting in voxel dimensions of 1.65 mm$^3$. This protocol also includes normalization, randoms smoothing, scatter correction, arc correction and a post reconstruction $3\times 3\times 3$ Gaussian filter. All data sets included matching X-ray CT scans used for attenuation correction. 

\begin{figure}[htbp]
	\centering
	\includegraphics[width=\columnwidth]{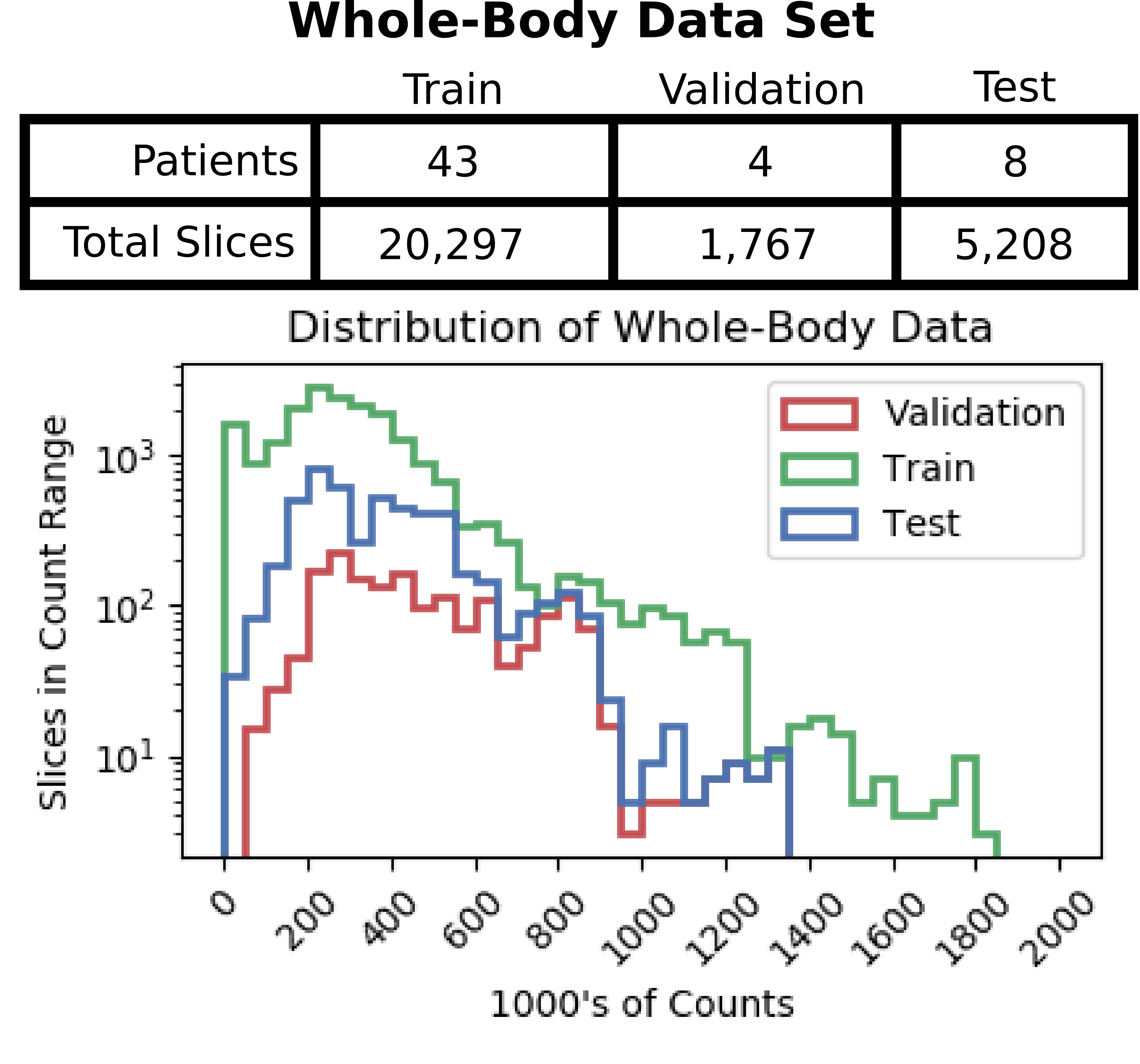}
	   \caption{The PET whole-body data set of 55 patients is split into training, validation and test subsets. A histogram of the distribution of counts shown on a logarithmic scale highlights how the majority of whole-body slices contain less than 800k counts.}
	\label{whole-body}
\end{figure}

Three different data sets were used to train and evaluate the performance of the FastPET neural network. The first is a whole-body data set consisting of 55 list-mode patient studies acquired with a continuous bed motion protocol using a flurodeoxyglucose ($^{18}$FDG) tracer and the OSEM+PSF reconstruction protocol with 3 iterations and 5 subsets. In this data set, each region of the body was in the PET field-of-view for approximately 2-3 minutes. The data set contains 27,272 total slices with a minimum axial depth of 435 slices up to a maximum of 1081 slices with an average axial depth of 496 slices. Figure~\ref{whole-body} details the separation into training, validation and test subsets along with a histogram of the overall distribution of the data set in counts per slice. At the outset 43 patients were designated for training, 4 patients designated for validation used to tune network parameters during development and the remaining 8 patients were strictly used for testing the final network. Generally speaking, PET whole-body data is challenging for neural networks due to the wide variety of anatomical structures and the wide spectrum of low count (i.e., high noise) regions as illustrated in Figure~\ref{whole-body} by the majority of slices in the data set containing 800k counts or less. 

The second data set is a simulated low-dose version of the whole-body data set discussed in the previous paragraph. The original data was decimated prior to creating the histo-images by randomly removing counts from the whole-body list-mode files with a probability of 0.75. However, the target images in the original data set were retained creating pairs of low-count histo-images and normal-count reconstructed targets. This enabled an investigation of the ability of FastPET to produce high quality images in a high-noise or low-dose situation. The split into training, validation and test subsets remained the same as for the normal-count data set.  

\begin{figure}[htbp]
	\centering
	\includegraphics[width=\columnwidth]{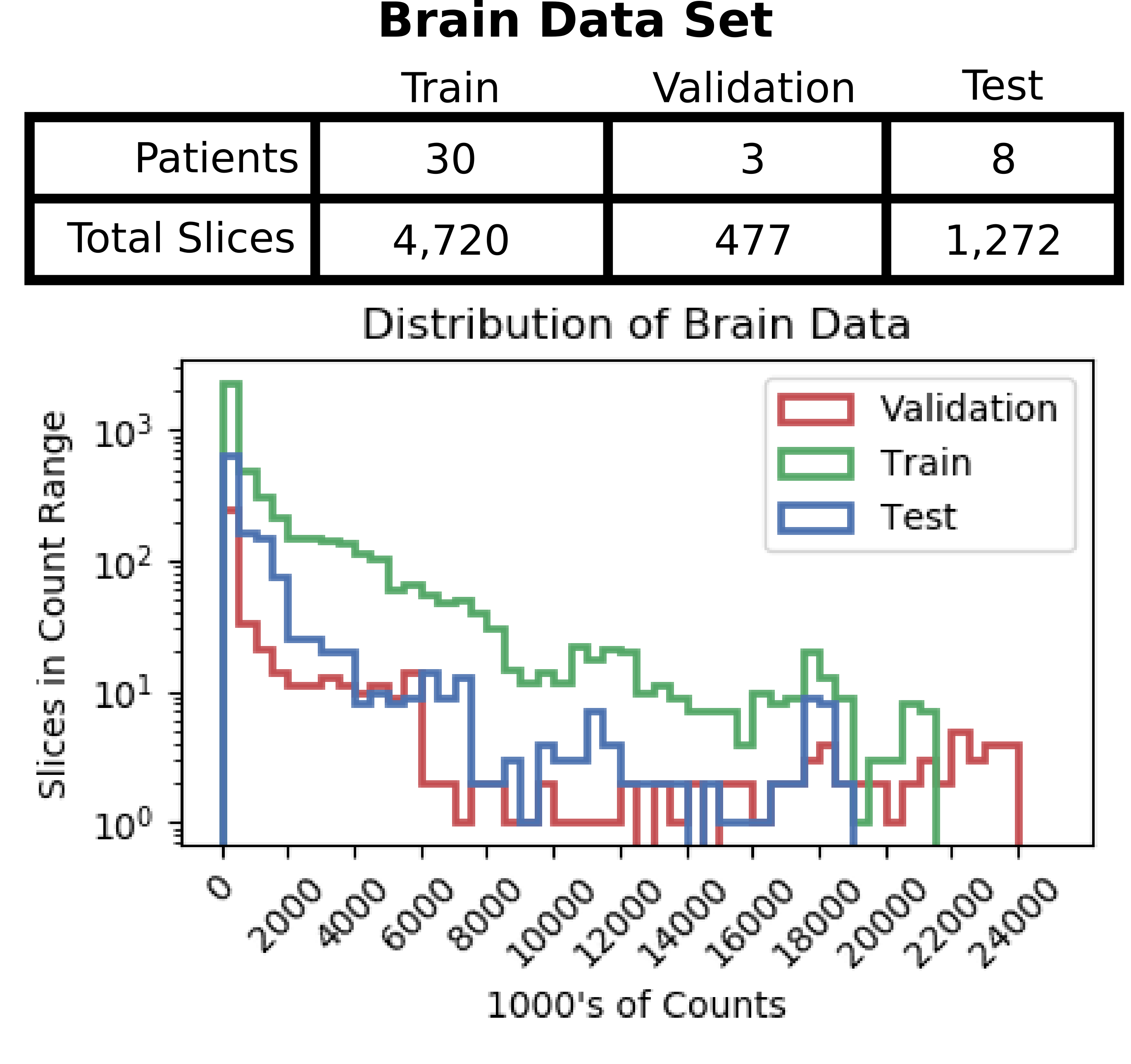}
	   \caption{The PET brain data set contains 41 patient studies that utilized a variety of tracers and acquisition times including flurodeoxyglucose (300 sec and 900 sec), fluorodopa (360 sec) and fluoroethyl-l-tyrosine (3,000 sec). A histogram of count distribution shown on a logarithmic scale illustrates the wide range of counts per slice in the data set.}
	\label{brain-data}
\end{figure}

The third data set consists of 41 single field-of-view neurology studies that vary in acquisition duration and the specific tracer used. Flurodeoxyglucose ($^{18}$FDG) was used on 4 patients with a 300 second acquisition and on 21 patients with a 900 second acquisition. Fluorodopa was used on 10 patients with a 360 second acquisition time and fluoroethyl-l-tyrosine ($^{18}$F-FET) was used on 6 patients with a 3,000 second acquisition time. With a consistent 159 slice field-of-view in each study, this provided a total of 6,519 slices allocated into training, validation and test subsets as shown in Figure~\ref{brain-data}. The patients in the test set were selected to include 5 with flurodeoxyglucose, including 2 with a 300 second acquisition, 2 with fluorodopa and 1 with fluoroethyl-l-tyrosine. This data set enables the evaluation of FastPET on a specific anatomical region with multiple tracers and with a wide distribution of count totals per slice as shown by the histogram in Figure~\ref{brain-data} where counts per slice range from nearly empty all the way to over 20 million counts in a single slice.   

\subsection{Neural Network Training}

The FastPET neural network was implemented with version 1.5 of the PyTorch\cite{paszke:2017} deep learning library and was trained, validated and tested on both Nvidia Titan RTX GPUs and a Supermicro HGX-2 with 16 Nvidia V100 GPUs. In scenarios where less GPU resources are available, smaller 3D chunks can be reconstructed and combined using weighted averaging or some other volume stitching method. FastPET is relatively insensitive to 3D chunk size allowing the adjustment of these dimensions without the need for retraining. FastPET training occurred over 500 epochs with each epoch drawing 2,048 samples from the training data set in batches ranging from 4 when using a single Titan RTX up to batches of 32 on the HGX-2. Each sample contained a tuple of histo-image, attenuation map and target image with an axial depth of 96 slices to match our available GPU capacity. The Adam optimizer\cite{Kingma:14}, which is similar to traditional stochastic gradient descent but additionally maintains a separate adaptive learning rate for each network parameter, was used with $\beta_1\!=\!0.5$ and $\beta_2\!=\!0.999$. In addition to the optimizer, a cyclic learning rate scheduler\cite{Smith:15} was employed to cycle the learning rate between a lower bound of 0.000005 and upper bound of 0.000045 with the amplitude of the cycle decaying exponentially over time towards the lower bound. This type of scheduler aids training by periodically raising the learning rate providing an opportunity for the network to escape sub-optimal local minimum and traverse saddle points more rapidly. 

\begin{figure}[h]
	\centering
	\includegraphics[width=0.7\columnwidth]{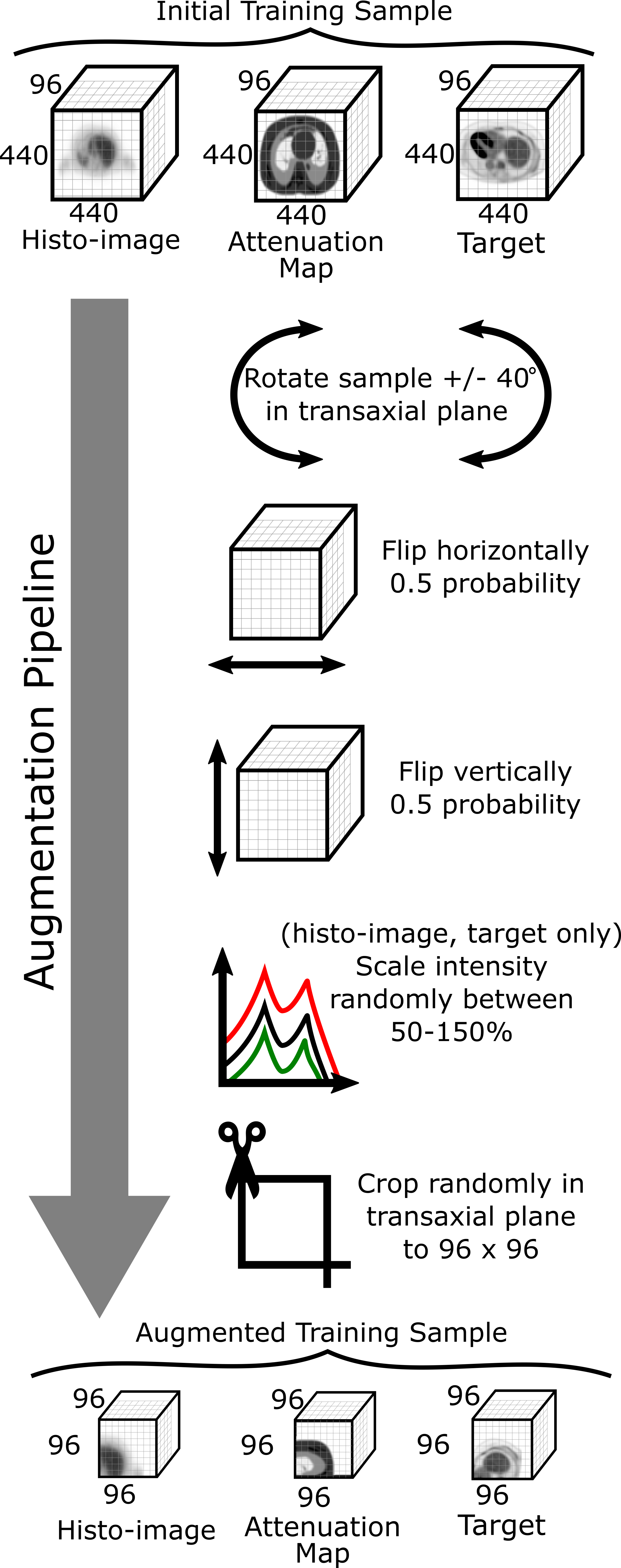}
	   \caption{The augmentation pipeline is utilized every time a sample is taken from the training data set. The continuous augmentation during training helps prevent overfitting and improves generalization. }
	\label{aug-pipeline}
\end{figure}

The training data is continuously augmented so the neural network never sees the same batch of training input, which helps prevent overfitting and improves network generalization when faced with a relatively limited data set. After randomly sampling a three-tuple of histo-image, attenuation map and target image from the training data subset, it is sent through the augmentation pipeline shown in Figure~\ref{aug-pipeline} and then added to the batch for the forward pass of the network. The samples progress through a series of augmentation steps each with a random component. These steps include rotation $\pm $ 40 degrees, horizontal and vertical flipping, intensity scaling from 50\%-150\% (only the histo-image and target), and cropping in the transaxial plane to $96\times 96$.   

During training, the optimizer minimizes the loss function through gradient descent. Zhao et al.\@ published research\cite{Zhao:17} on network loss functions for image generation and repair suggesting a weighted combination of an absolute measure, such as element-wise L1 loss, and a perceptual measure, such as multi-scale structural similarity (MS-SSIM) \cite{wang:04}, was optimal. The FastPET loss function trades-off these two elements using a dynamically balanced scale factor. The complete loss function used to minimize the difference between the proposed method's reconstructed image $\hat{x}$ and a target image $x$ was thus made to consist of two terms, specifically:
\begin{align}
\mathcal{L}(\hat{x},x) = (1-\alpha) \ \text{MAE}(\hat{x},x) + \alpha \ \text{MS-SSIM}(\hat{x},x)
\label{loss_fun}
\end{align}

The mean absolute error (MAE) denotes the common L1 voxel loss between reconstructed image $\hat{x}$ and target image $x$ calculated over all voxels, and the MS-SSIM term uses the standard formulation of structural similarity \cite{wang:04} over 5 image scales with a window size of $11\times 11$. 

A dynamically calculated $\alpha$ value balances the MAE and MS-SSIM losses against one another.
That is:
\begin{align}
   \alpha=\frac{\sum\limits_{j=i}^{i+n-1}\text{MAE}_j}{ \sum\limits_{j=i}^{i+n-1}\text{MAE}_j+\sum\limits_{j=i}^{i+n-1}\text{MS-SSIM}_j}
\end{align}
where $i$ and $j$ are iteration steps and $n$ denotes the width of a running average window.

\begin{figure}[htbp]
	\centering
	\includegraphics[width=0.85\columnwidth]{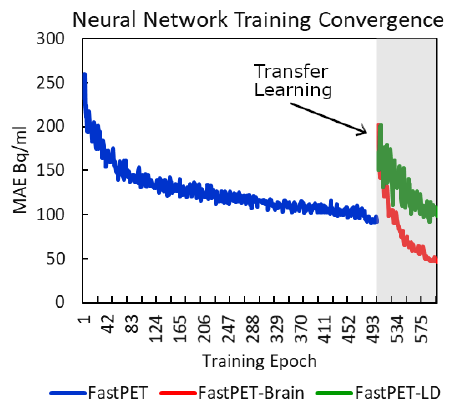}
	   \caption{The baseline FastPET network trained on whole-body data was used as the baseline to perform transfer learning for two additional networks adapted to low-dose data and neurology applications. }
	\label{convergence}
\end{figure}

A separate neural network was trained for each of the three data sets using a transfer learning approach illustrated in the convergence chart in Figure~\ref{convergence}. The baseline network, simply named FastPET, was trained from scratch on the normal count whole-body data set for 500 epochs. Then the training for each of the remaining two data sets started with the baseline network and was trained for an additional 100 epochs. The network trained on the simulated low-dose whole-body data set is designated as FastPET-LD, while the network trained on the neurology data set is termed FastPET-Brain.  

\section{Experiments}
In this section, we study the performance of the proposed method on the whole-body and the low-dose whole-body data sets by measuring structural similarity, mean absolute error and examining reconstructed images. We examine additional quantitative measures through the use of two physical phantoms. We then move on to FastPET-Brain reconstructions fine tuned on a neurology data set containing three different tracers. Finally, we report on the reconstruction speed of the proposed method compared to conventional approaches.   
 
\subsection{Whole-body and Whole-body Low-dose}
The proposed architecture was trained for 500 epochs on the whole-body data set to create the FastPET network. This baseline network was then trained for an additional 100 epochs using transfer learning on the simulated low-dose whole-body data set with 75\% of the counts removed to create the FastPET-LD network. Note that in all cases in this section where FastPET-LD is mentioned, the raw data has been decimated to 25\% counts including training, validation, test, and phantom data. Both of these network variants reconstructed the subset of 8 test patients from their respective data sets. A summary of MS-SSIM and MAE measures across these test sets is shown in Table~\ref{ssim_comparison} and Table~\ref{mae_comparison}.
 
The MS-SSIM values in Table~\ref{ssim_comparison} perceptually measure how similar the reconstructed images are to their target images within a range from 0 to 1 where values closer to 1 indicate more similarity. The value shown in the table is the average value calculated across all slices of a particular patient. The FastPET and target images were normalized by dividing each pixel by the max pixel value found in the union of the two images. This provided a bounded range of pixel values, which is important in calculating structural similarity. The FastPET reconstructions on whole-body data are relatively consistent across the test set with a minimum value of 0.962 and a maximum value of 0.989. The FastPET-LD reconstructions on the low-dose version are likewise relatively consistent. Of significant note is the close performance between the normal count network with an average value of 0.981 and the low-dose network having only a slightly degraded performance with an average value of 0.978.    

\begin{table}[h]
\caption{Average multi-scale structural similarity measures across the patient test set for both normal count and low-dose whole-body networks and additionally the histo-images to serve as a third reference point. The close similarity in FastPET and FastPET-LD performance shows that removing 75\% of counts from the histo-images only slightly reduces perceptual performance.}
\label{ssim_comparison}
\centering
\begin{tabular}{|c|c|c|c|}
\hline 
\textbf{Test Patient}        & \textbf{Histo-Image} & \textbf{FastPET} & \textbf{FastPET-LD} \\ \hline
Patient 1   &  0.656  &  0.982    &  0.980   \\ \hline 
Patient 2   &  0.545  &  0.989    &  0.987   \\ \hline
Patient 3   &  0.553  &  0.981    &  0.978   \\ \hline
Patient 4   &  0.593  &  0.987    &  0.985   \\ \hline      
Patient 5   &  0.679  &  0.962    &  0.957   \\ \hline 
Patient 6   &  0.567  &  0.981    &  0.978   \\ \hline
Patient 7   &  0.569  &  0.983    &  0.980   \\ \hline
Patient 8   &  0.559  &  0.980    &  0.976   \\ \hline 
Average     &  0.590  &  0.981    &  0.978   \\ \hline
\end{tabular}
\end{table}

Table~\ref{mae_comparison} shows the MAE values of the reconstructed images versus the target images in Bq/ml. Again, we note a relatively consistent performance across the test set for both networks with FastPET achieving a minimum MAE of 19.83~Bq/ml and a maximum MAE of 47.75~Bq/ml. Similar to structural similarity, FastPET-LD achieves very similar values with only a slight degradation in performance.  

\begin{table}[h]
\caption{Average mean absolute error measured across the patient test set for both normal count and low-dose whole-body FastPET networks. Similar to the perceptual measure, both networks are relatively consistent across the test set. Notably, the low-dose network exhibits nearly the same performance as the normal count network. }
\label{mae_comparison}
\centering
\begin{tabular}{|c|c|c|c|} 
\hline 
\textbf{Test Patient}  & \textbf{FastPET} (Bq/ml) & \textbf{FastPET-LD} (Bq/ml) \\ \hline
Patient 1    &  19.83    &  21.35   \\ \hline 
Patient 2    &  28.79    &  31.00   \\ \hline
Patient 3    &  47.75    &  51.40   \\ \hline
Patient 4    &  26.05    &  28.05   \\ \hline      
Patient 5    &  27.47    &  29.58   \\ \hline 
Patient 6    &  27.13    &  29.22   \\ \hline
Patient 7    &  23.37    &  25.16   \\ \hline
Patient 8    &  34.13    &  36.71   \\ \hline 
Average      &  29.32    &  31.56   \\ \hline
\end{tabular}
\end{table}

\begin{figure}[htbp]
	\centering
	\includegraphics[width=0.95\columnwidth]{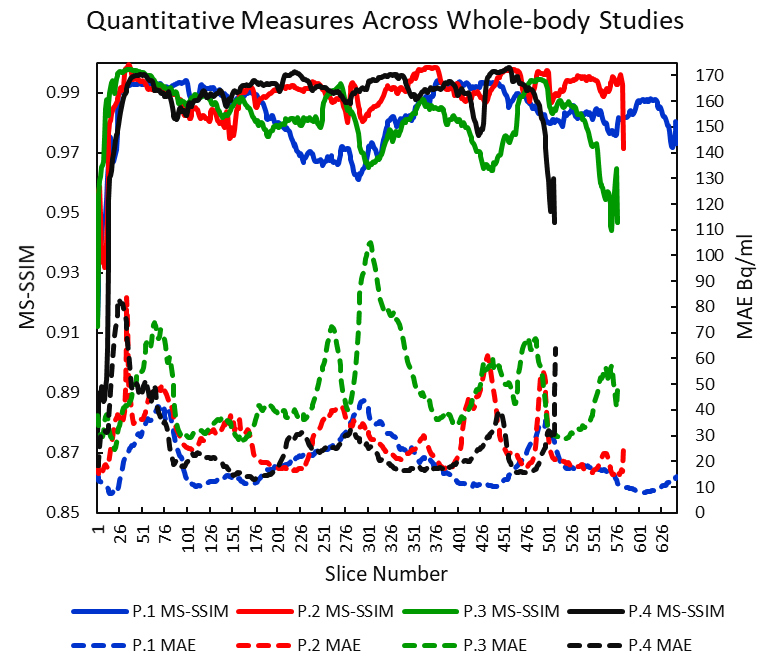}
	   \caption{Perceptual and absolute measures relative to the target images for FastPET are shown for individual slices for the first four patients in the test data set illustrating the variance across a whole-body study. }
	\label{wb_quant}
\end{figure}

While the average absolute and perceptual measures are shown in the preceding tables, the variance in performance across a patient study is illustrated in Figure~\ref{wb_quant} where the slice-by-slice measures for MS-SSIM (left axis; solid line) and MAE (right axis; dotted line) are shown. Only the first four test patients in the whole-body test set are displayed to preserve clarity in the chart. Note how both the perceptual and absolute measures vary across the axial span. In the evaluation of specific transaxial images in the following sections, slice 306 and 575 from test patient 3, which contain some of the largest single slice perceptual and absolute errors, are examined. The variance of image slices across the whole-body axial extent containing a wide range of structures and count levels should be investigated in future work to see if better consistency is achievable.  

\begin{figure*}[htbp]
	\centering
	\includegraphics[width=\textwidth]{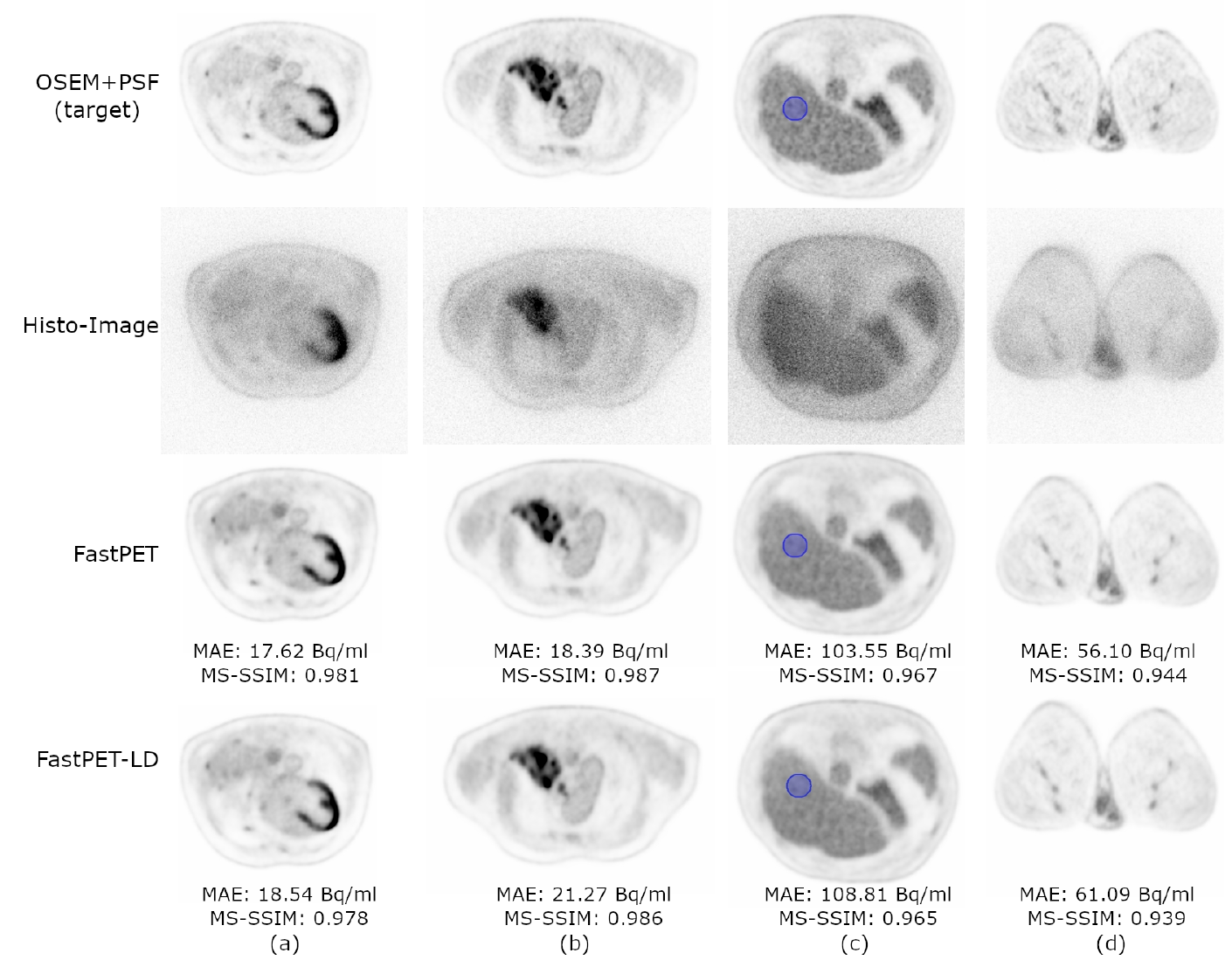}
	   \caption{Transaxial slices showing the OSEM+PSF target image, the histo-image input and both the FastPET whole-body reconstruction and the FastPET-LD low-dose reconstruction from only 25\% counts. Columns (c) and (d) were specifically chosen as slices of relatively high error amoung the test set.}
	\label{transaxial}
\end{figure*}

A variety of transaxial slices are provided in Figure~\ref{transaxial} showing the OSEM+PSF (target) image, histo-image (input to the neural network) and both reconstructions from the whole-body data set using FastPET and from the low-dose data set using FastPET-LD. Column (a) shows tracer in the walls of the heart and column (b) illustrates a cluster of lesions in the upper chest. In both cases the neural network images closely resemble the target image, but do contain less of the high frequency texture and have an overall more smoothed appearance than the iterative reconstruction. Column (c) in the figure is slice 306 from test patient 3 referenced previously in the context of Figure~\ref{wb_quant}, and has relatively high error compared to the other slices in the study. A visual examination of these images finds that most of the structural detail is indeed present in the neural network images, but possibly contains an overall intensity bias. This is confirmed by analyzing a region of interest in the liver highlighted in blue where the mean voxel value for the OSEM+PSF image is 6,147 Bq/ml, while the mean values for FastPET and FastPET-LD are 4,911 Bq/ml and 4,719 Bq/ml respectively. In column (d) slice 575 from test patient 3 is examined, which is an end slice of the study anatomically located in the upper thigh. This image exhibits the characteristic falloff in absolute and perceptual performance seen in FastPET reconstructions in the outer tomographic slices. Similar to column (c), most of the structural detail is present, but again seems to suffer from an overall intensity bias, which in this case is confirmed by taking the mean voxel value for the entire image where the mean of OSEM+PSF is 767 Bq/ml and FastPET and FastPET-LD have mean values of 534 Bq/ml and 483 Bq/ml.   

\begin{figure*}[htbp]
	\centering
	\includegraphics[width=\textwidth]{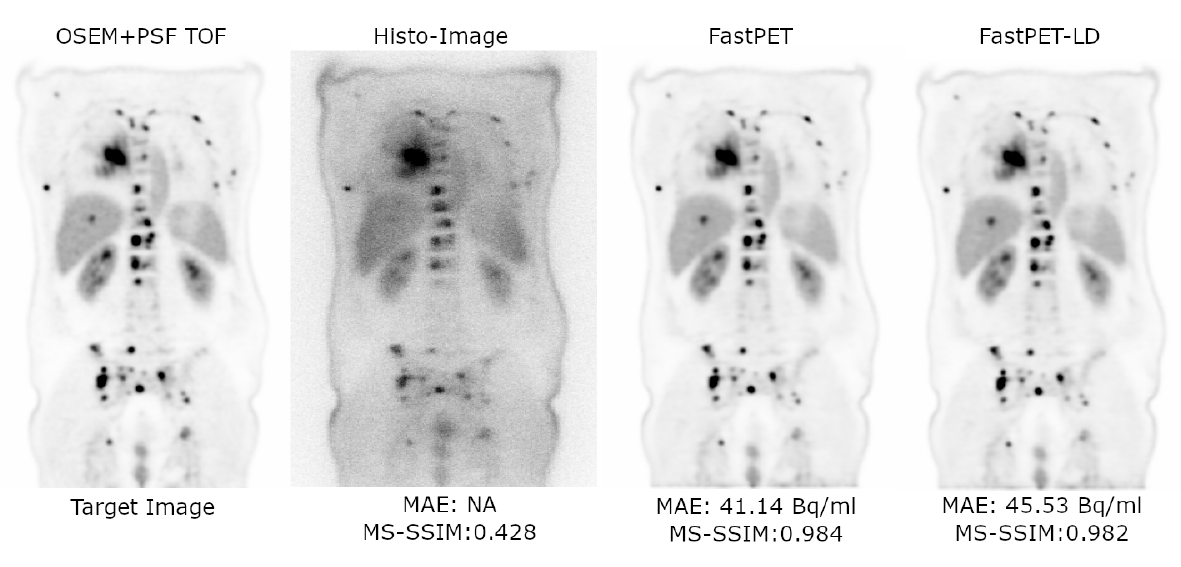}
	   \caption{A coronal view comparing conventional OSEM+PSF TOF reconstruction, the FastPET neural network and the low-dose FastPET-LD variant is shown for test patient 8 along with the histo-image input to the network. The mean absolute error and multi-scale structural similarity is also displayed below each image. }
	\label{coronal}
\end{figure*}

Coronal views of a whole-body study containing many lesions with varying characteristics are shown in Figure~\ref{coronal}. This example is a challenging test for neural network reconstruction due to the varying size, location, intensity and background activity of the lesions and the fact that it provides a view across the tomographic slices in the study. While the neural network reconstructions contain less high frequency noise and structure, as previously noted, they do contain all of the many lesions in the target image.  

\begin{figure}
    \centering
    \includegraphics[width=0.70\columnwidth]{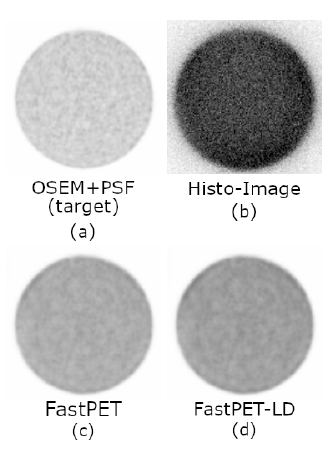}
    \caption{Uniform phantom reconstructions for OSEM+PSF, histo-image and FastPET and FastPET-LD networks illustrating smoothness and additional bias exhibited by the neural network reconstruction.}
    \label{uniform}
\end{figure}

To further examine the bias and smoothness of neural network images, a uniform phantom and a phantom with varying sized spheres both containing FDG$^{18}$ were analyzed. The uniform phantom is shown in Figure~\ref{uniform} with target, input and FastPET and FastPET-LD reconstructions. Visually the neural network reconstructions are smoother and darker than the OSEM+PSF image. The statistics from a spherical region of interest (not shown in the figure) with a diameter of 80 mm placed in the center of the uniform phantom are shown in Table~\ref{uniformity}. In this case the neural network images are positively biased with mean voxel values of 6,137 Bq/ml for FastPET and 6,197 Bq/ml FastPET-LD compared to the target image of 5,761 Bq/ml. The smoothness of the neural network reconstructions is also confirmed with significantly lower voxel standard deviation and minimum and maximum values closer to their mean than OSEM+PSF.  

\begin{table}[h]
\caption{Statistics from an 80 mm spherical region of interest centered in the uniform phantom shown in Figure~\ref{uniform} highlighting the smoothness and some bias exhibited by FastPET reconstruction}
\label{uniformity}
\centering
\begin{tabular}{|c|c|c|c|c|}
\hline 
\textbf{Method}  & \textbf{Mean} & \textbf{Std Dev} & \textbf{Min} & \textbf{Max} \\ \hline
OSEM+PSF    &  5,761    &  827 & 3,217 & 9,764   \\ \hline 
FastPET     &  6,137    &  562 & 4,392 & 8,561   \\ \hline
FastPET-LD  &  6,197    &  572 & 4,463 & 8,515   \\ \hline
\end{tabular}
\end{table}

\begin{figure}
    \centering
    \includegraphics[width=0.85\columnwidth]{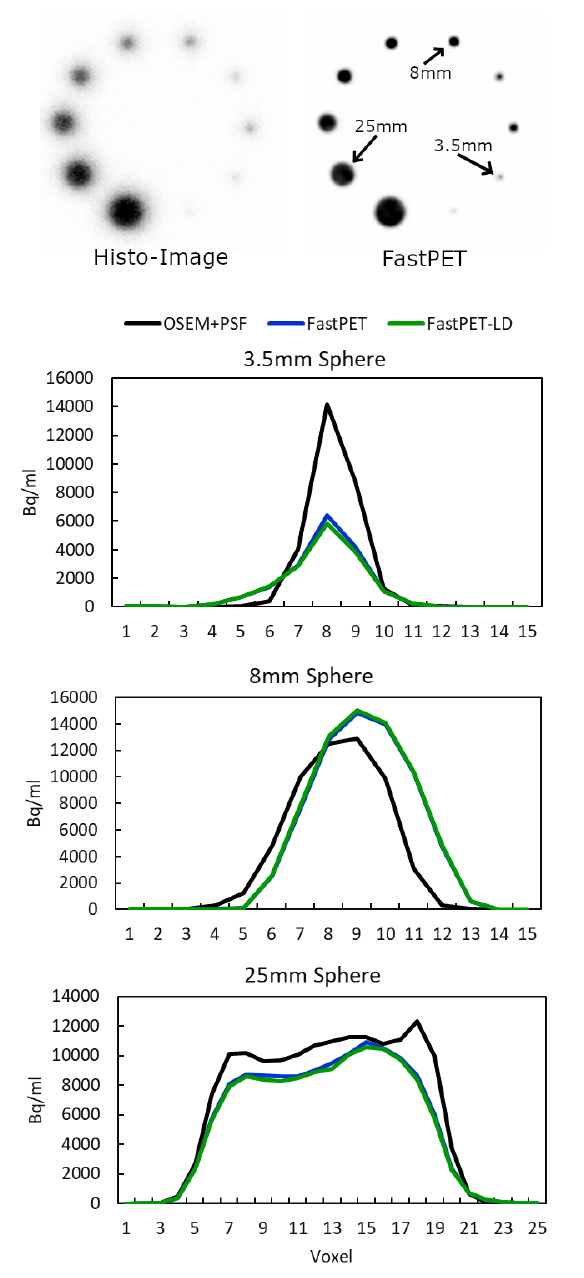}
    \caption{Line profiles for 3.5 mm, 8 mm and 25mm spheres from a physical phantom are shown for OSEM+PSF, FastPET and FastPET-LD reconstructions.}
    \label{line_profile}
\end{figure}

Using a physical phantom with 10 different sized spheres, line profiles shown in Figure~\ref{line_profile} were generated from three of the spheres with varying diameters of 3.5~mm, 8~mm and 25~mm. Additionally, Table~\ref{spheres_table} contains the full width at half maximum (FWHM) measurements for each of the line profiles. For the smallest 3.5~mm sphere, both the neural network line profiles demonstrate a smoothing of the object with FWHM measurements of 3.93~mm and 4.19~mm and a significantly lower overall magnitude. The OSEM+PSF reconstruction by contrast performs better at recovering tracer in this small region with a higher peak value and FWHM of 2.98~mm. In the mid-sized 8~mm sphere the line profiles are similar in magnitude and width with FWHM measurements of 6.32~mm, 6.89~mm and 6.93~mm for OSEM+PSF, FastPET and FastPET-LD. In the largest 25~mm sphere again the FWHM measures are similar across the three reconstruction methods, but the lower response to high frequencies is apparent in how the neural network profiles have dampened transitions from peak values. In contrast OSEM+PSF exhibits sharper transitions. This behavior is consistent with images shown in Figures~\ref{transaxial} and \ref{coronal} where the transitions from tracer concentration to background are sharper in the OSEM+PSF images compared to FastPET and FastPET-LD.

Overall the FastPET reconstruction results on whole-body data indicate the potential of the neural network to produce high quality outputs from histo-images. While there is still significant progress to be made on reducing bias and increasing quantitative accuracy, the surprisingly close performance of the neural network on low-dose data points to the possibility of even higher quality reconstructions given an improved training data set and a better data normalization strategy. 

\begin{table}[h]
\caption{The full width at half maximum measurements for the line profiles in Figure~\ref{line_profile} for each of the three sphere sizes. }
\label{spheres_table}
\centering
\begin{tabular}{|c|c|c|c|}
\hline 
\textbf{Method}  & \textbf{3.5mm} & \textbf{8mm} & \textbf{25mm}\\ \hline
OSEM+PSF    &  2.98 mm    &  6.32 mm & 20.76 mm    \\ \hline 
FastPET     &  3.93 mm    &  6.89 mm & 19.72 mm    \\ \hline
FastPET-LD  &  4.19 mm    &  6.93 mm & 19.62 mm    \\ \hline
\end{tabular}
\end{table}

\subsection{Neurology Application Focused Network}

\begin{figure*}[htbp]
	\centering
	\includegraphics[width=0.85\textwidth]{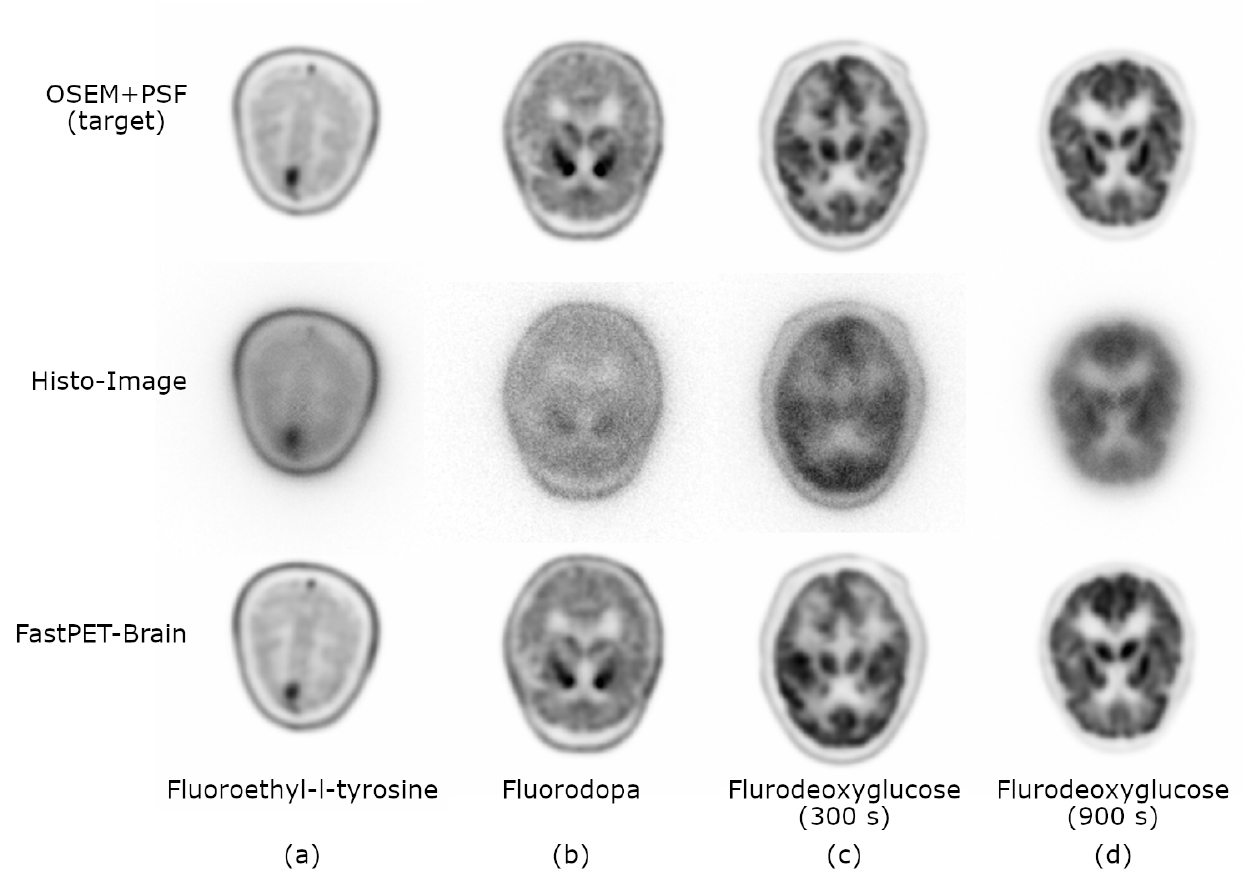}
	   \caption{A selection of transaxial slices from the neurology test set with each of the different type of tracers and acquisition duration. The results on this limited data set indicate a strong possibility that neural network reconstruction can generalize across different PET tracers. }
	\label{brain}
\end{figure*}

Starting with the trained FastPET network as a baseline, an additional 100 epochs of training were conducted using a neurology data set containing three different tracers resulting in the FastPET-Brain network. Narrowing the reconstruction application to only images of the head allows the neural network to specialize in reconstructing an anatomically smaller distribution of images. However, one open question in the domain of direct neural network reconstruction that flows in the opposite direction of specialization is can a neural network generalize to multiple tracers. This section contains the first known experiment in PET imaging that begins to examine whether a neural network can generalize in this way. 

Table~\ref{brain_comparison} shows a summary of the reconstruction results for the 8 patients in the neurology test set. The first column contains the name of the tracer and the duration of the acquisition. The second column contains the total number of counts in the study and the third and fourth columns are measures of MAE and MS-SSIM. The number of counts in a study varies widely from only 32.4 million counts in a fluorodopa study to over 1.1 billion counts in a 900 second $^{18}$FDG study. The two lower count fluorodopa studies have the lowest MAE error with values of 8.29 Bq/ml and 6.53 Bq/ml while also maintaining high perceptual scores of 0.990 and 0.989. The 3,000 second fluoroethyl-l-tyrosine study with over 600 million counts maintains a proportionally low MAE of 17.35 Bq/ml and high perceptual score of 0.991. The $^{18}$FDG studies show mixed results with the 300 second studies having relatively high MAE values proportional to the total counts and MS-SSIM scores below the test set average while two of the three 900 second studies have proportionally low MAE scores and also perceptual scores above the test set average. 

\begin{table}[h]
\caption{Mean absolute error and multi-scale structural similarity measured across the neurology test set for FastPET-Brain reconstructions.}
\label{brain_comparison}
\centering
\begin{tabular}{|c|c|c|c|}
\hline 
\textbf{Patient,Tracer,Time} & \textbf{Study Counts} &\textbf{MAE} (Bq/ml) & \textbf{MS-SSIM} \\ \hline
P1, F-dopa, 360 s       &  40.9 M     & 8.29    & 0.990   \\ \hline 
P2, $^{18}$FDG 300 s    &  113.2 M    & 28.13   & 0.984   \\ \hline
P3, $^{18}$FDG 300 s    &  111.5 M    & 28.14   & 0.979  \\ \hline
P4, F-dopa, 360 s       &  32.4 M     & 6.53    & 0.989  \\ \hline      
P5, $^{18}$FDG 900 s    &  233.8 M    & 40.85   & 0.977   \\ \hline 
P6, $^{18}$F-FET 3,000 s &  602.9 M    & 17.35   & 0.991 \\ \hline
P7, $^{18}$FDG 900 s    &  406.0 M    & 18.40    & 0.989 \\ \hline
P8, $^{18}$FDG 900 s    &  1112.8 M   & 30.97   & 0.989 \\ \hline 
Average                 &  373.2 M    & 22.33   & 0.986   \\ \hline
\end{tabular}
\end{table}

Sample images, network inputs and targets from the neurology test set containing each of the tracer types are shown in Figure~\ref{brain}. The different ways the various tracers uptake in the brain is apparent in the images shown. This scenario is notable since the neural network is not able to strictly learn the anatomical structures of the brain and must learn a more general distribution of the positron emitters. Additionally, the perceptual performance outperforms the more general whole-body network and the MAE performance is also better despite the majority of the neurology data set having a higher average tracer concentration. Although this experiment used an admittedly limited data set, the results do indicate the benefit of narrowing the trained distribution and also the possibility of training reconstruction neural networks to generalize across different PET tracers.

\subsection{Reconstruction Speed}

\begin{figure}[ht]
	\centering
	\includegraphics[width=0.85\columnwidth]{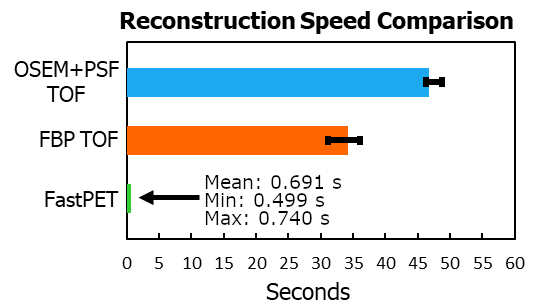}
	   \caption{Comparison of reconstruction speeds using Biograph Vision PET/CT data shows that FastPET is about 67x faster than OSEM and 50x faster than FBP using relatively modest hardware. }
	\label{speed_comparison}
\end{figure}

The simplicity of the MLAP histogrammer combined with the computational efficiency of the neural network provides the area where FastPET excels past conventional methods, namely, reconstruction speed. A comparison of reconstruction speed between FastPET and Filtered Back-Projection (FBP) and OSEM+PSF both with TOF is shown in Figure~\ref{speed_comparison}. The conventional methods start from an uncompressed sinogram while FastPET starts from a histo-image. These two starting points are roughly computationally equivalent since both cases have the raw coincidence events histogrammed into a data structure appropriate for the method. The timing results shown are for reconstructing a $159x440x440$ image volume using Biograph Vision PET/CT data. The conventional reconstructions were performed on a HP Z8G4 computer containing an Intel Xeon Gold 6154 CPU running at 3.0 GHz. While 36 CPU cores are available on this system, the reconstruction was restricted to 12 by Siemens factory settings. The iterative OSEM+PSF reconstruction was based on 3 iterations and 5 subsets, and both iterative and analytical methods included attenuation and scatter correction as well as post-reconstruction smoothing. FastPET reconstructions were performed on an HP z840 computer with 2 Intel E5-2630 CPUs each with 10 cores running at 2.2 GHz and a single Nvidia Titan RTX GPU with 24 GB of memory. 

The results demonstrate the efficiency of neural network reconstruction with FastPET taking an average of 0.7 seconds and 19.4 GB of GPU memory to reconstruct the $159x440x440$ image volume. By comparison, OSEM+PSF reconstruction with TOF took an average of 46.7 seconds. The same reconstruction averaged 34.2 seconds for the FBP algorithm. While this may not significantly affect the clinical workflow when a single static scan is considered, protocols that  require many reconstructions would benefit from FastPET. In cardiology scans for example, the study of perfusion often requires around 10 gated reconstructions, while the study of blood flow often requires around 26 dynamic frames. In these scenarios FastPET would complete these reconstructions in respectively 7 seconds and 18 seconds compared to about 8 and 20 minutes for OSEM+PSF. Additionally, the potential for entirely new reconstruction applications are enabled by ultra fast reconstruction such as interventional procedures, improved motion correction with sub-second frames, or a near real-time PET image viewer.

\section{Discussion and Challenges}
Overall the FastPET method of reconstructing histo-images directly with a neural network has shown significant potential to produce quality 3D image volumes with high efficiency. However, during our research and experimentation a number of key challenges were encountered that warrant further consideration and investigation. With the three primary components of neural network reconstruction being the network architecture, training methods and the data set, we found the majority of the noted challenges lay in the latter two categories and that time spent adjusting network hyperparameters often did not return significant dividends. 

While we assert that FastPET produces high quality PET images, certain aspects could be improved. The neural network images are smoother and lower noise than the OSEM+PSF targets. While a high signal-to-noise-ratio is usually viewed positively, in FastPET it comes with a loss in sharpness at the border between high tracer concentration and the background and is simply the failure of the network to learn the highest frequency content in the target image. It is possible that a network with more capacity could improve this, but it is likely better addressed through improved training methods such as more sophisticated loss functions that penalize the absence of high frequency details.    

Another challenge is dealing with a non-systematic bias evident in many images caused by the lack of a good normalization scheme for the input and target data. Although we experimented with a number of normalization techniques, we ultimately opted to instead let the neural network adapt to the differences in data ranges in the input and output data. This decision was based on a desire to maintain quantitative values and avoid having to choose an arbitrary scalar for voxel values that are theoretically not bounded by a defined range. Finding a solution to this problem such as normalizing the data based on SUV\cite{luyao:2019} will be a focus of future research.

By far the biggest challenge associated with direct neural network reconstruction is having sufficient amounts of good data available. The quality of the reconstruction and the ability to generalize to unseen data is dominated by the quality and quantity of the training data. The few tens of thousands of tomographic slices used here are by no means representative of the entire real distribution of possible PET images containing a wide range of patient sizes, scatter profiles, disease and radioactive tracer types. Additionally, using OSEM+PSF reconstructions as training targets introduces a sub-optimal approximation of the ground truth. The solution may be to create computer models, like Monte Carlo simulation, that can generate unlimited data closely following the physics of positron decay and the system matrix of a particular scanner.   

On the positive side, FastPET has been shown to consistently produce high quality 3D image volumes across a wide range of data without loss of significant structural details. FastPET accomplishes this with near real-time speed while requiring significantly fewer network parameters than other direct reconstruction methods. Worthy of note, the same network fine tuned on a low-dose data set with only 25\% counts performed nearly identically to the full-count neural network. Additionally, the FastPET-Brain network variant showed promise at generalizing across multiple tracers when trained on a neurology specific data set. 

Looking toward future work there are a number of important areas to explore. First is the applicability of this technique to scanners with lower timing resolution in the 400 ps to 700 ps range, which encompasses the majority of scanners in current use. It is also worthwhile to examine the importance of including the attenuation maps in the reconstruction and the impact of potential misalignment with the PET data. Another area of future research is developing more practical applications for the proposed technique such as aiding in motion correction, or developing a scan protocol with a real-time feedback loop. Lastly and most important is addressing the challenging problem of the absence and perhaps the impractical nature of curating large comprehensive PET data sets for training that include the raw event data. One solution would be to develop better simulation models making the need to collect clinical data unnecessary and another option could train the network on a baseline data set and then continue to learn the specific data distributions at a clinical site once deployed. 

\section{Conclusion}
 FastPET has been shown capable of true 3D image reconstruction in near real-time. Raw list-mode PET data is processed by a most likely annihilation position histogrammer. The resulting histo-images are input to a 3D convolutional neural network along with the corresponding attenuation maps to produce PET images that are quantitatively and qualitatively similar to conventional iterative time-of-flight reconstruction targets. Task specific networks can be trained using the process of transfer learning, e.g., for low dose PET imaging and anatomically focused brain imaging with multiple tracers. Areas of improvement include increasing high frequency details and bias reduction.

\section*{Acknowledgment}
The authors would like to thank Paul Luk for providing valuable support and feedback throughout the development of FastPET.

\ifCLASSOPTIONcaptionsoff
  \newpage
\fi



%

\bibliography{fastPET}
\bibliographystyle{IEEEtran}

%

\vskip -2\baselineskip plus -1fil
\begin{IEEEbiographynophoto}{Dr. William Whiteley} is a research scientist at Siemens Medical Solutions USA, Inc. He received his BSEE degree from Vanderbilt University, Nashville, TN, his MSEE degree from Stanford University, Stanford, CA and his PhD from The University of Tennessee, Knoxville. His research interests include artificial intelligence, deep learning, medical imaging, and data mining.
\end{IEEEbiographynophoto}

\vskip -2\baselineskip plus -1fil

\begin{IEEEbiographynophoto}{Dr. Vladimir Panin} is a Senior Key Expert at Siemens Medical Solutions USA, Inc. He received his Diploma in Theoretical Physics from Moscow Engineering Physics Institute and PhD from Physics Department, University of Utah. His research interests include PET/CT image reconstruction and data corrections. 
\end{IEEEbiographynophoto}
\vskip -2\baselineskip plus -1fil
\begin{IEEEbiographynophoto}{Dr. Chuanyu Zhou} is a staff software engineer at Siemens Medical Solutions USA, Inc. He received his PhD degree from the department of Aerospace Engineering and MS degree from department of Electrical and Computer Engineering department, Iowa State University. His research interests include PET/CT image reconstruction, image processing, deep learning and high performance computing. 
\end{IEEEbiographynophoto}
\vskip -2\baselineskip plus -1fil
\begin{IEEEbiographynophoto}{Dr. Jorge Cabello} is a research scientist in the Physics and Reconstruction group at Siemens Medical Solutions USA, Inc. He received his PhD from Surrey University, UK, and his MSEE and BSEE degrees from Universidad de Alcala, Spain. His research interests cover image reconstruction, Monte Carlo simulations, artificial intelligence and detector physics.
\end{IEEEbiographynophoto}
\vskip -2\baselineskip plus -1fil
\begin{IEEEbiographynophoto}{Dr. Deepak K Bharkhada} is a staff Scientist in the molecular imaging division of Siemens Medical Solutions USA, Inc. He has a PhD in Biomedical Engineering from a joint program between Wake Forest University and Virginia Polytechnic, MS from Drexel University, and BE from Mumbai University. His research interests include image reconstruction and corrections for tomographic imaging modalities like PET and CT and artificial intelligence.
\end{IEEEbiographynophoto}
\vskip -2\baselineskip plus -1fil
\begin{IEEEbiographynophoto}{Dr. Jens Gregor} received his PhD in Electrical Engineering from Aalborg University, Denmark, in 1991. He then joined the faculty at The University of Tennessee, Knoxville, where he currently holds the position of Professor and Associate Department Head in the Department of Electrical Engineering and Computer Science. His research has covered different imaging modalities including X-ray and neutron CT, SPECT, and PET with applications ranging from industrial and security imaging to preclinical and clinical imaging.
\end{IEEEbiographynophoto}




\end{document}